\documentstyle[prl,aps,floats,epsf,color]{revtex}
\begin{document}
\baselineskip=12pt
\def\be{\begin{equation}}
\def\ee{\end{equation}}
\def\bea{\begin{eqnarray}}
\def\eea{\end{eqnarray}}
\def\E{{\rm e}}
\def\bearst{\begin{eqnarray*}}
\def\eearst{\end{eqnarray*}}
\def\peleven{\parbox{11cm}}
\def\peffec{\peight{\bearst\eearst}\hfill\peleven}
\def\pspace{\peight{\bearst\eearst}\hfill}
\def\ptwelve{\parbox{12cm}}
\def\peight{\parbox{8mm}}
\twocolumn
[\hsize\textwidth\columnwidth\hsize\csname@twocolumnfalse\endcsname

\title
{  NEW COMPUTATIONAL APPROACHES TO ANALYSIS OF INTERBEAT INTERVALS
IN HUMAN SUBJECTS}

\author
{  M. Reza Rahimi Tabar,$^{1,2}$ Fatemeh Ghasemi,$^3$ J.
Peinke,$^4$ R. Friedrich,$^5$ Kamran Kaviani,$^6$ F.
Taghavi,$^1$\\
S. Sadeghi,$^1$ G. Bijani,$^1$ and Muhammad Sahimi$^7$}

\vskip 1cm

\address
{$^1$Department of Physics, Sharif University of Technology, P.O.
Box
11365-9161, Tehran 11365, Iran\\
$^2$CNRS UMR 6529, Observatoire de la C$\hat o$te d'Azur, BP 4229,
06304 Nice Cedex 4, France\\
$^3$Institute for Studies in Theoretical Physics and Mathematics,
P.O. Box 19395-5531, Tehran 19395, Iran \\
$^4$Carl von Ossietzky University, Institute of Physics, D-26111
Oldenburg, Germany\\
$^5$Institute for Theoretical Physics, University of M\"unster,
D-48149 M\"unster, Germany\\
$^6$ Department of Physics, Az-zahra University, P.O. Box 19834,
Tehran 19834,
Iran\\
$^7$Mork Family Department of Chemical Engineering and Materials
Science, University of Southern California, Los Angeles,
California 90089-1211, USA}
 \maketitle


\hspace{.3in}
\newpage
]

\noindent{\bf 1. Introduction}

Complex, self- regulating systems such
as the human heart must process inputs with a broad range of characteristics
to change physiological data and time series.$^{1–3}$ Many of these physiological time series
seem to be highly chaotic, represent nonstationary data,
and fluctuate in an irregular and complex manner. One hypothesis
is that the seemingly chaotic structure of physiological
time series arises from external and intrinsic
perturbations that push the system away from a homeostatic
set point. An alternative hypothesis is that the fluctuations
are due, at least in part, to the system's underlying dynamics.
In this review, we describe new computational approaches—
based on new theoretical concepts—for analyzing
physiological time series. We'll show that the application
of these methods could potentially lead to a novel diagnostic
tool for distinguishing healthy individuals from those
with congestive heart failure (CHF).

{\bf Physiological Time Series}

Recent research suggests that physiological time series can
possess fractal and self-similar properties, which are characterized
by the existence of long-range correlations (with the
correlation function being a power-law type). However, until
recently, the analysis of such fluctuations' fractal properties
was restricted to computing certain characteristics based on
the second moment of the data, such as the power spectrum
and the two-point autocorrelation function. These analyses
indicated that the fractal behavior of healthy, free-running
physiological systems could be characterized, at least in some
cases, by 1/f-like scaling of the power spectra over a wide range
of time scales.$^{4,5}$ A time series that exhibits such long-range
correlations with a power-law correlation function and is also
homogeneous (different parts of the series have identical statistical
properties) is called a monofractal series.
However, many physiological time series are inhomogeneous
in the sense that distinct statistical and scaling properties
characterize different parts of the series. In addition,
there is some evidence that physiological dynamics can exhibit
nonlinear properties.$^{6–-12}$ Such features are often associated
with multifractal behavior—the presence of
long-range power-law correlations in the higher moments
of the time series—which, unlike monofractals, are nonlinear
functions of the second moment's scaling exponents.$^{13}$
Up until recently, though, robust demonstration
of multifractality of nonstationary time series was hampered
by problems related to significant bias in the estimates
of the data's singularity spectrum, due to the time
series' diverging negative moments. A new wavelet-based
multifractal formalism helps address such problems.$^{13}$
Among physiological time series, the study of the statistical
properties of heartbeat interval sequences has attracted
much attention.$^{14–-17}$ The interest is at least partly due to the
facts that
• the heart rate is under direct neuroautonomic control;
• interbeat interval variability is readily measured by noninvasive
means; and
• analysis of heart-rate dynamics could provide important
diagnostic and prognostic information.
Thus, extensive analysis of interbeat interval variability represents
an important quantity for elucidating possibly nonhomeostatic
physiological variability.
Figure 1 shows examples of cardiac interbeat time series
(the output of a spatially and temporally integrated neuroautonomic
control system) for healthy individuals and
those with CHF. In the conventional approaches to analyzing
such data, we would assume the apparent noise has no
meaningful structure, so we wouldn't expect to gain any
understanding of the underlying system through the study of
such fluctuations. Conventional studies that focus on averaged
quantities therefore usually ignore these fluctuations—
in fact, they're often labeled as noise to distinguish them
from the true time series of interest.
However, by adapting and extending methods developed
in modern statistical physics and nonlinear dynamics, the
physiological fluctuations in Figure 1 can be shown to exhibit
an unexpected hidden scaling structure.$^{5,10,13,18,19}$
Moreover, the fluctuation dynamics and associated scaling
features can change with pathological perturbation. These
discoveries have raised the possibility that understanding the
origin of such temporal structures and their alterations
through disease could elucidate certain basic aspects of
heart-rate control mechanisms and increase the potential for
clinical monitoring.
But despite this considerable progress, several interesting
features must still be analyzed and interpreted. The theoretical
concepts we discuss here are based on the possible
Markov properties of the time series; a cascade of information
from large time scales to small ones that are built based
on increments in the time series; and the extended self-similar
properties of the beat-to-beat fluctuations of healthy
subjects as well as those with CHF. The method we describe
uses a set of data for a given phenomenon that contains
a degree of stochasticity and numerically constructs a
relatively simple equation that governs the phenomenon.
In addition to being accurate, this method is quite general,
can provide a rational explanation for complex features of
the phenomenon under study, and requires no scaling feature
or assumption.
As we analyze cardiac interbeat intervals, we'll also look
at new methods for computing the Kramers-Moyal (KM)
coefficients for the increments of interbeat intervals fluctuations,$\Delta
x(\tau)=[x(t+\tau)-x(t)]/\sigma_\tau$, where $\sigma_\tau$ is the
standard deviations of the increments in the interbeats data. Whereas the
first and second KM coefficients (representing the drift and
diffusion coefficients in a Fokker-Planck [FP] equation) have
well-defined values, the third- and fourth-order KM coeffi-
cients might be small. If so, we can numerically construct an
FP evolution equation for the probability density function,
$P(\Delta x,\tau)$ which, in turn,  can be used to gain information on
the PDF's evolution as a function of the time scale
${\tau}.^{20–23}$

\bigskip
\noindent{\bf Regeneration of Stochastic Processes }

\bigskip

Let's start by examining the computations that lead to the
numerical construction of a stochastic equation. This equation
describes the phenomenon that generates the data set,
which is then utilized to reconstruct the original time series.
Two basic steps are involved in the numerical analysis of the
data and their reconstruction.
Data Examination
We must first examine the data to see whether they follow a
Markov chain and, if so, we estimate the Markov time scale
$t_M$. As is well known, a given process with a degree of stochasticity
can have a finite or an infinite Markov time scale,
which is the minimum time interval over which the data can
be considered to be a Markov process.$^{20,24}$
To determine the Markov scale $t_M$, we
note that a complete characterization of the statistical
properties of stochastic fluctuations of a quantity $x(t)$
requires the numerical evaluation of the joint PDF
$P_n(x_1,t_1;\cdots;x_n,t_n)$ for an arbitrary $n$, the number of
the data points in the time series $x(t)$. If the time series
$x(t)$ is a Markov process, an important simplification can be
made as $P_n$, the $n$-point joint PDF, is generated by the {\it
product} of the conditional probabilities
$P(x_{i+1},t_{i+1}|x_i,t_i)$, for $i=1,\cdots,n-1$. A necessary
condition for $x(t)$ to be a Markov process is that the
Chapman-Kolmogorov (CK) equation,
\begin{equation}
P(x_2,t_2|x_1,t_1)=\int\hbox{d}(x_3)\;P(x_2,t_2|x_3,t_3)\;P(x_3,t_3|x_1,t_1)\;,
\end{equation}
should hold for any value of $t_3$ in the interval $t_2<t_3<t_1$.
One should check the validity of the CK equation for various $x_1$
by comparing the directly-computed conditional probability
distributions $P(x_2,t_2|x_1,t_1)$ with the ones computed
according to right side of Eq. (1). The simplest way to determine
$t_M$ for stationary or homogeneous data is the numerical
computation of the quantity,
$S=|P(x_2,t_2|x_1,t_1)-\int\hbox{d}x_3
P(x_2,t_2|x_3,t_3)\,P(x_3,t_3|x_1,t_1)|$, for given $x_1$ and
$x_2$, in terms of, for example, $t_3-t_1$ (taking into account
the possible numerical errors in estimating $S$). Then,
$t_M=t_3-t_1$ for that value of $t_3-t_1$ for which $S$ vanishes
or is nearly zero (achieves a minimum).
\begin{figure}
\epsfxsize=7truecm\epsfbox{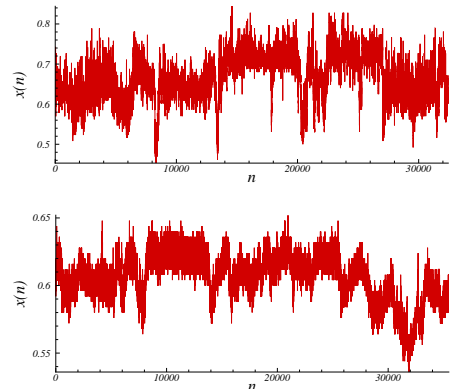}
 \narrowtext \caption{ Interbeats fluctuations of healthy subjects
(top), and those with congestive heart failure (bottom).}
 \end{figure}
\begin{figure}
\epsfxsize=7truecm\epsfbox{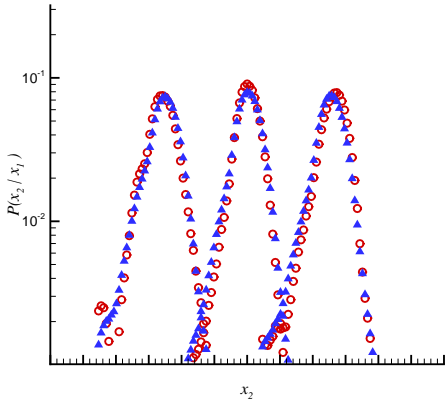} \narrowtext \caption{ Test of
Chapman-Kolmogorov equation for $x_1=-5$, $x_1=0$ and $x_1=5$. The
bold and open symbols represent, respectively, the
directly-evaluated PDF and the integrated PDF. The PDFs are
shifted in the vertical directions for better presentation. Values
of $x$ are measured in units of the standard deviation. }
 \end{figure}
(2) Numerical construction of an effective stochastic equation
that describes the fluctuations of the quantity $x(t)$,
representing the time series, constitutes the second step. The CK
equation yields an evolution equation for the PDF $P(x,t)$ across
the scales $t$. The CK equation, when formulated in differential
form, yields a master equation which takes the form of a FP
equation:
\begin{equation}
\frac{d}{dt}P(x,t)=\left[-\frac{\partial}{\partial x}D^{(1)}(x,t)+
\frac{\partial^2}{\partial x^2}D^{(2)}(x,t)\right]P(x,t)\;.
\end{equation}
The drift and diffusion coefficients, $D^{(1)}(x,t)$ and
$D^{(2)}(x,t)$, are computed directly from the data and the
moments $M^{(k)}$ of the conditional probability distributions:
\begin{equation}
D^{(k)}(x,t)=\frac{1}{k!}\lim_{\Delta t\to 0}M^{(k)}\;,
\end{equation}
\begin{equation}
M^{(k)}=\frac{1}{\Delta t}\int dx'(x'-x)^k P(x',t+\Delta t|x,t)\;.
\end{equation}
Note that the above FP formulation is equivalent to the following
Langevin equation:
\begin{equation}
\frac{d}{dt}x(t)=D^{(1)}(x)+ \sqrt{D^{(2)}(x)}\;\; f(t)\;,
\end{equation}
where $f(t)$ is a {\it random force} with zero mean and Gaussian
statistics, $\delta$-correlated in $t$, i.e., $\langle
f(t)f(t')\rangle=2\delta(t-t')$. We note that the numerical
reconstruction of a stochastic process does {\it not} imply that
the data do not contain any correlations, or that the above
formulation ignores the correlations.

Equation (5) enables us to {\it reconstruct} a stochastic time
series $x(t)$, which is similar to the original one {\it in the
statistical sense}. The stochastic process $x(t)$ is regenerated
by iterating Eq. (5) which yields a series of data {\it without
memory}. To compare the regenerated series with the original
$x(t)$, we must take the temporal interval in the numerical
discretization of Eq. (5) to be unity (or renormalize it to unity,
if need be). However, the Markov time is typically greater than
unity. Therefore, we correlate the data over the Markov time scale
$t_M$, for which there are a number of methods.$^{21,22,25}$ A new
technique that we have used in our own studies, which we refer to
as the {\it kernel} method, is one according to which one
considers a kernel function $K(u)$ that satisfies the condition
that,
\begin{equation}
\int_{-\infty}^\infty K(u)du=1\;,
\end{equation}
such that the data are determined, or reconstructed, by
\begin{equation}
x(t)=\frac{1}{nh}\sum_{i=1}^nx(t_i)K\left(\frac{t-t_i}{h}\right)\;,
\end{equation}
where $h$ is the window width. For example, one of the most
accurate kernels is the standard normal density function,
$K(u)=(2\pi)^{-1/2}\exp(-\frac{1}{2}u^2)$. In essence, the kernel
method represents the the time series as a sum of ``bumps'' placed
at the ``observation'' points, with its kernel determining the
shape of the bumps, and its window width $h$ fixing their width.
It is evident that, over the scale $h$, the kernel method
correlates the data.

\begin{figure}
\epsfxsize=7truecm\epsfbox{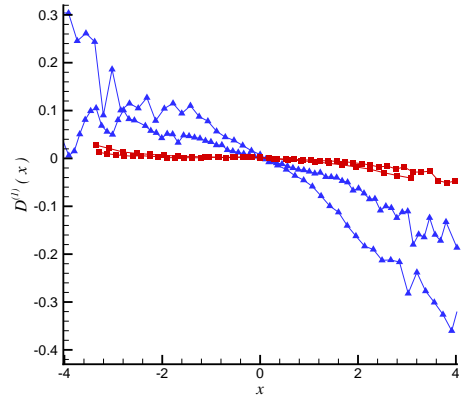}
\epsfxsize=7truecm\epsfbox{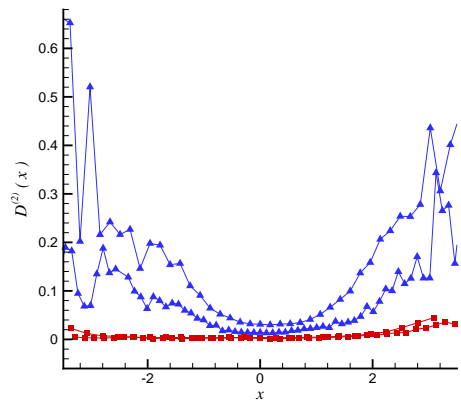} \narrowtext \caption{ The drift
and diffusion coefficients $D^{(1)}(x)$ and $D^{(2)}(x)$,
estimated by Eq. (3). For the healthy subjects (triangles)
$D^{(1)}(x)$ and $D^{(2)}(x)$ follow linear and quadratic behavior
in $x$, while for patients with CHF (squares) they follow third-
and fourth-order behavior in $x$.}
 \end{figure}
\begin{figure}
\epsfxsize=7truecm\epsfbox{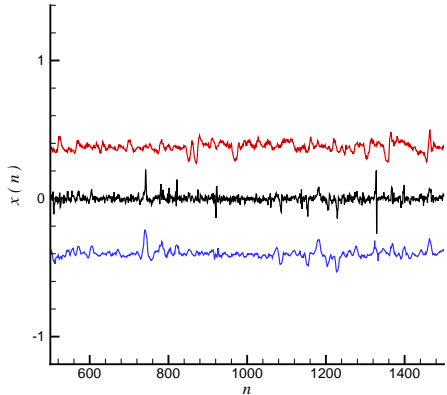} \narrowtext \caption{ The
curves show, from top to bottom, the actual interbeat data (for a
healthy subject), the regenerated data using the corresponding
Langevin equation, and the regenerated data using the kernel
method. The time series are shifted in the vertical directions for
better presentation.}
 \end{figure}
\bigskip
\noindent{\bf 3. Analysis of Fluctuations in Human Heartbeats}

\bigskip
To show how the above reconstruction method is used in practice,
and to demonstrate its utility, the method has been applied to
reconstruction of the fluctuations in human heartbeats of both
healthy and ill subjects by taking $h\simeq t_M$. Recent
studies$^{5,19,26,27}$ reveal that under normal conditions,
beat-to-beat fluctuations in the heart rates might display
extended correlations of the type typically exhibited by dynamical
systems far from equilibrium. It has been argued,$^{26}$ for
example, that the various stages of sleep might be characterized
by extended correlations of heart rates separated by a large
number of beats. While the existence of extended correlations in
the fluctuations of human heartbeats is an interesting and
important result, we argue that the Markov time scale $t_M$ and
the associated drift and diffusion coefficients of the interbeat
fluctuations of healthy subjects and those with CHF help one to
better distinguish the two classes of subjects, particularly at
the early stages of the disease, as these quantities have
completely different behaviour for the two classes of subjects.

Both daytime (12:00 pm to 18:00 pm) and nighttime (12:00 am to
6:00 am) heartbeat time series of healthy subjects, and the
daytime records of patients with CHF have been analyzed by this
method.$^{45,46}$ The data base includes 10 healthy subjects (7
females and 3 males with ages between 20 and 50, and average age
of 34.3 years), and 12 subjects with CHF (3 females and 9 males
with ages between 22 and 71, and average age of 60.8 years).
Figure 1 presents the typical data.

As the first step, the Markov time scale $t_M$ of the data is
computed. From the daytime data for healthy subjects the values of
$t_M$ are computed to be (all the values are measured in units of
the average time scale for the beat-to-beat times of each
subject), $t_M=3,3,3,1,2,3,3,2,3$ and 2. The corresponding values
for the nighttime records are, $t_M$ are $3,3,1,3,3,2,3,3,2$ and
$3$, respectively, comparable to those for the daytime. On the
other hand, for the daytime records of the patients with CHF, the
computed Markov time scales are,
$t_M=151,258,760,542,231,257,864,8,366,393,385$, and 276.
Therefore, the healthy subjects are characterized by $t_M$ values
that are much smaller than those of the patients with CHF. Thus,
one has an unambiguous quantity for distinguishing the two classes
of patients.

Next, the validity of the CK equation for describing the
phenomenon is checked for several $x_1$-triplets by comparing the
directly-computed conditional probability distributions
$P(x_2,t_2|x_1,t_1)$ with the ones computed according to right
side of Eq. (1). Here, $x$ is the interbeat and for all the
samples we define, $x\equiv(x-\bar{x})/\sigma$, where $\bar x$ and
$\sigma$ are the mean and standard deviations of the interbeat
data. In Figure 2, the two PDFs, computed by the two methods, are
compared. Assuming the statistical errors to be the square root of
the number of events in each bin, the two PDFs are {\it
statistically} identical.

The corresponding drift and diffusion coefficients, $D^{(1)}(x)$
and $D^{(2)}(x)$, are displayed in Figure 3, demonstrating that,
in addition to the Markov time scale $t_M$, the two coefficients
provide another important indicator for distinguishing the ill
from healthy subjects: For the healthy subjects the drift
$D^{(1)}$ and the diffusion coefficients $D^{(2)}(x)$ follow
(approximately) linear and quadratic functions of $x$,
respectively, whereas the corresponding coefficients for patients
with CHF follow (approximately) third- and fourth-order equations
in $x$. Thus, for the healthy subjects,
\begin{equation}
D^{(1)}(x)=-0.12x\;,
\end{equation}
\begin{equation}
D^{(2)}(x)=(5-4.2x+7x^2)\times 10^{-2}\;,
\end{equation}
whereas for the patients with CHF,
\begin{equation}
D^{(1)}(x)=-(26x+18x^2+7x^3)\times 10^{-4}\;,
\end{equation}
\begin{equation}
D^{(2)}(x)=(6-7x+5x^2+3x^3+2x^4)\times 10^{-4}\;,
\end{equation}
which were also obtained by Kuusela.$^{29}$ For other data bases
measured for other patients, the functional dependence of
$D^{(1)}$ and $D^{(2)}(x)$ would be the same, but with different
numerical coefficients. The order of magnitude of the coefficients
would be the same for all the healthy subjects, and likewise for
those with CHF (see also Wolf {\it et al.}$^{20}$). Moreover, if
one analyzes different parts of the time series separately, one
finds, (1) practically the same Markov time scale for different
parts of the time series, but with some differences in the
numerical values of the drift and diffusion coefficients, and (2)
that the drift and diffusion coefficients for different parts of
the time series have the same {\it functional forms}, but with
{\it different coefficients} in equations such as (8)-(11). Hence,
one can distinguish the data for sleeping times from those
patients when they are awake.

There is yet another important difference between the heartbeat
dynamics of the two classes of subjects: Compared with the healthy
subjects, the drift and diffusion coefficients for the patients
with CHF are very small, reflecting, in some sense, their large
Markov time scale $t_M$. Large Markov times $t_M$ imply longer
correlation lengths for the data, and it is well-known that the
diffusivity in correlated system is {\it smaller} than those in
random ones. Hence, one may use the Markov time scales, and the
dependence of the drift and diffusion coefficients on $x$, as well
as their comparative magnitudes, for characterizing the dynamics
of human heartbeats and their fluctuations, and to distinguish
healthy subjects from those with CHF.

How accurate is the reconstruction method? Shown in Figure 4 is a
comparison between the original time series $x(n)$ and those
reconstructed by the Langevin equation [by, for example, using
Eqs. (5), (8) and (9)] and the kernel method. While both methods
generate series that look similar to the original data, the kernel
method appears to better mimic the behavior of the original data.
To demonstrate the accuracy of Eq. (7), Figure 5 compares the
second moment of the stochastic function,
$C_2(m)=\langle[x(0)-x(m)]^2 \rangle$, for both the measured and
reconstructed data using the kernel method. The agreement between
the two is excellent. However, it is well-known that such
agreement is not sufficient for proving the accuracy of a
reconstruction method. Hence, the accuracy of the reconstructed
higher-order structure function, $S_n=\langle
|x(t_1)-x(t_2)|^n\rangle,^{28,29}$ was also checked. It was found
that the agreement between $S_n$ for the original and
reconstructed time series for $n\leq 5$ is excellent, while the
difference between higher-order moments of the two times series,
which are related to the tails of the PDF of the $x-$increments,
increases.
\begin{figure}
\epsfxsize=7truecm\epsfbox{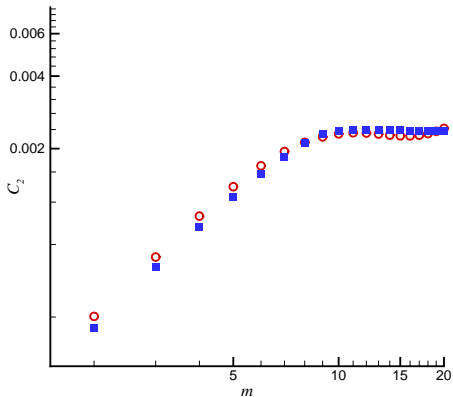} \narrowtext \caption{
Logarithmic plot of the second moment of the height-difference
versus $m$, for the actual data (circles) and the samples
regenerated by the kernel method (squares). The corresponding time
series are plotted in Fig. 4.}
 \end{figure}

\noindent{\bf The Cascade of Information from Large to Small Time
Scales}

We would argue that if long-range, {\it nondecaying}
correlations do exist in the time series, then one may not be able
to use the above reconstruction method for analyzing them because,
as is well-known, the correlations in a Markov process decay
exponentially. Aside from the fact that even in such cases the
method described above provides an unambiguous way of
distinguishing healthy subjects from those with CHF, which we
believe is more effective than simply analyzing the data to see
what type of correlations may exist in the data (see below), we
argue that the nondecaying correlations do not, in fact, pose any
limitations to the fundamental ideas and concepts of the
reconstruction method described above.

The reason is that, even if the above reconstruction method fails
to describe long-range, nondecaying correlations in the data, one
can still analyze the data based on the same method by invoking an
important result recently pointed out by several
groups.$^{20,21,23-25}$ They studied the evolution of the PDF of
several stochastic properties of turbulent free jets, and rough
surfaces. They pointed out that the conditional PDF of the {\it
increments}  of a stochastic field, such as the increments in the
velocity field in the turbulent flow or heights fluctuations of
rough surface, satisfies the CK equation even if the velocity
field or the height function itself contains long-range,
nondecaying correlations. This enabled them to derive a FP
equation for describing the systems under study. Hence, one has a
way of analyzing correlated stochastic time series or data in
terms of the corresponding FP and CK equations. We now describe
the conditions under which such a formulation can be utilized.
\begin{figure}
\epsfxsize=7truecm\epsfbox{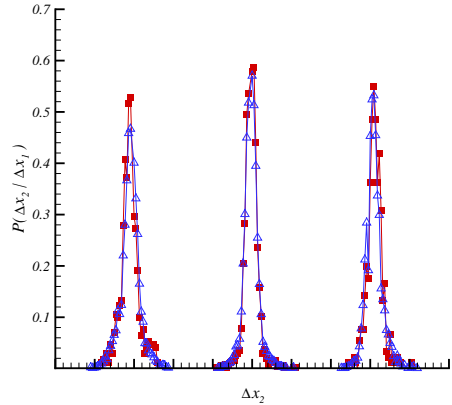} \narrowtext \caption{Test of
Chapman-Kolmogorov equation for $\Delta x_1= -0.42 $, $\Delta
x_1=0$ and $\Delta x_1=0.42$. The solid and open symbols
represent, respectively, the directly-evaluated PDF and the one
obtained from Eq. (1). The PDFs are shifted in the horizontal
directions for clarity. Values of $\Delta x$ are measured in units
of the standard deviation of the increments. The time scales
$\tau_1$, $\tau_2$ and $\tau_3$ are $10$, $30$, and $20$,
respectively.}
\end{figure}
\begin{figure}
\epsfxsize=6truecm\epsfbox{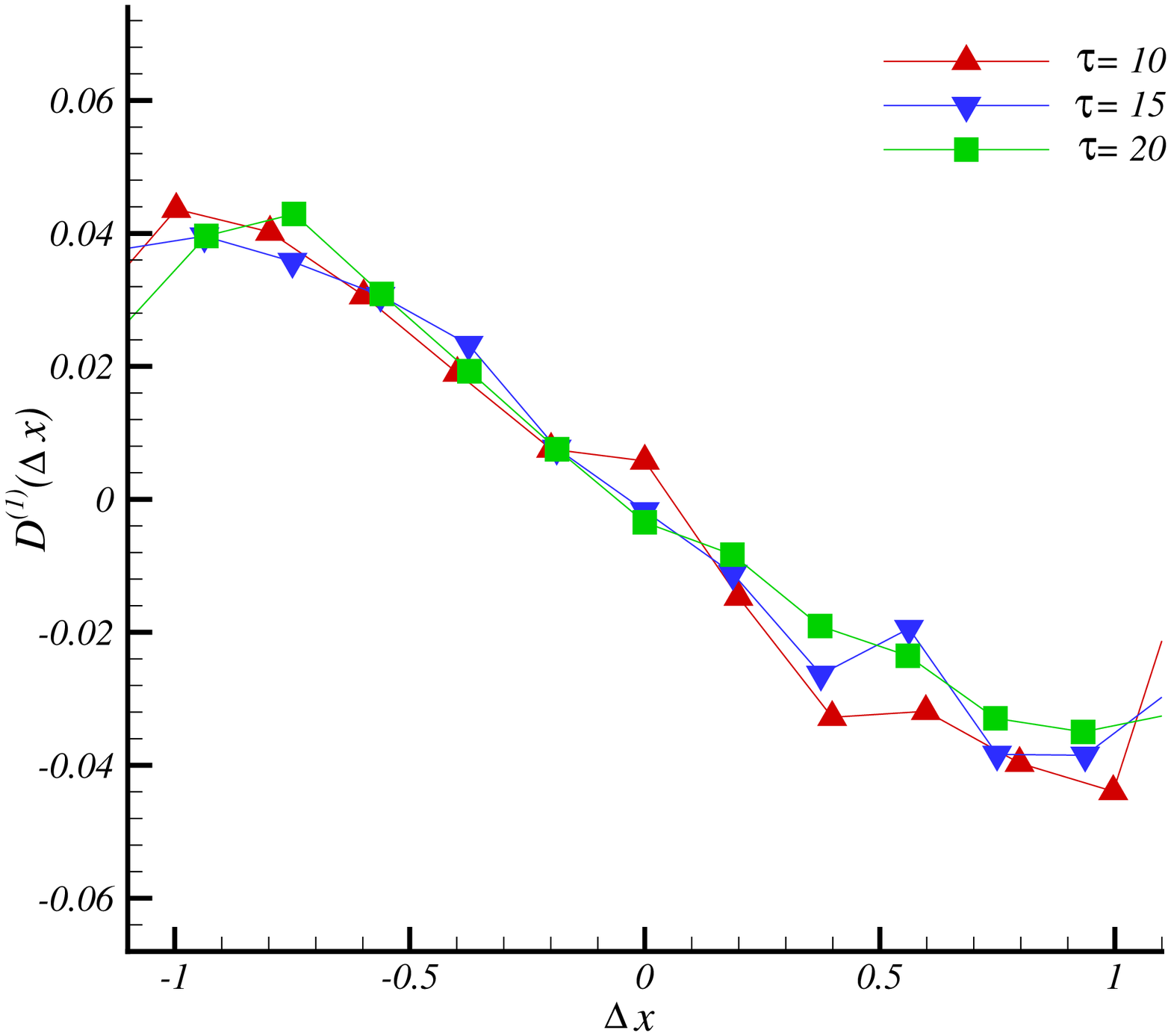}
\epsfxsize=6truecm\epsfbox{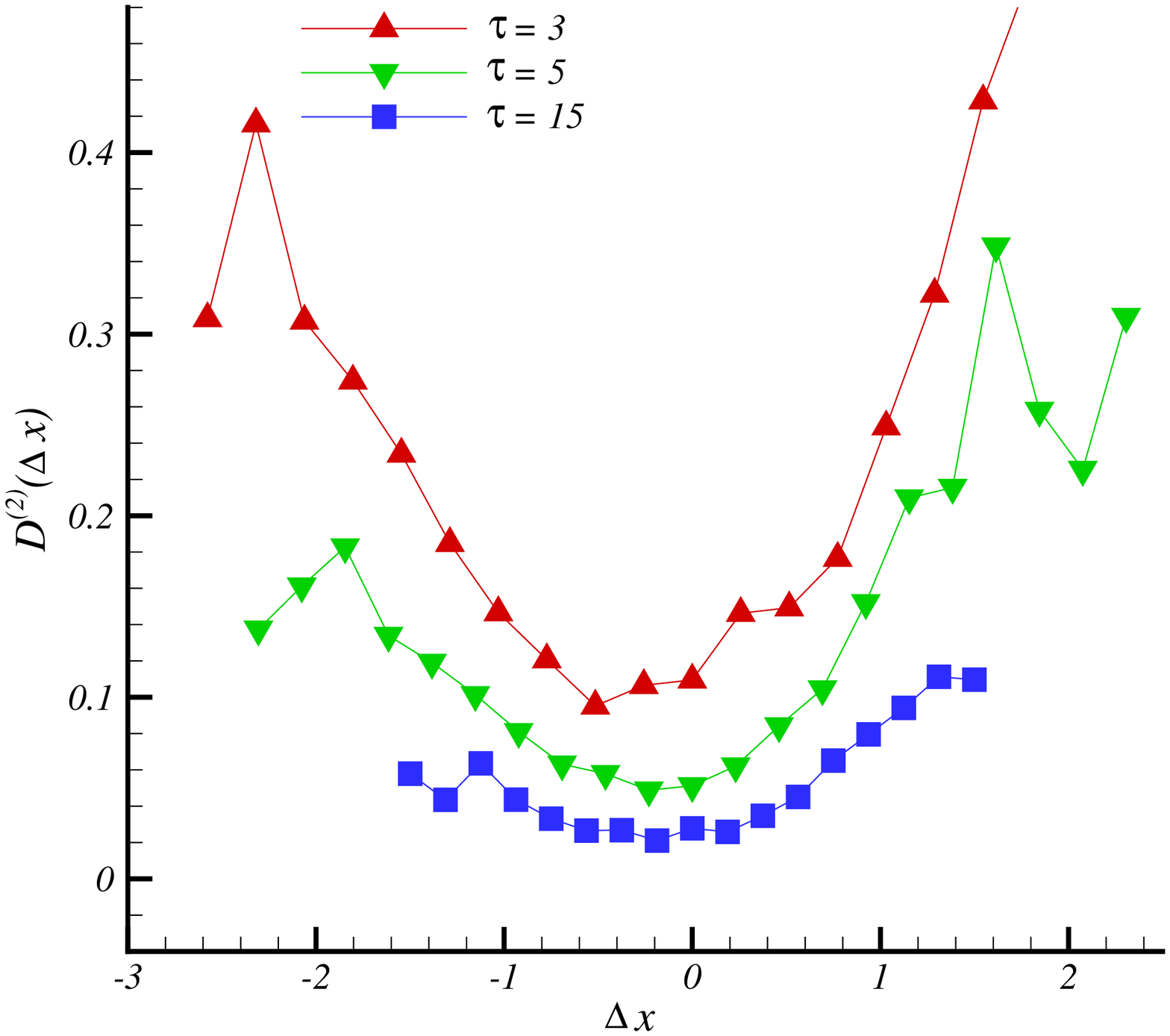}\narrowtext\caption{The drift and
diffusion coefficients $D^{(1)}(\Delta x)$ and $D^{(2)}(\Delta
x)$, estimated from Eq. (3) for a healthy subject, follow linear
and quadratic behavior, respectively.}
\end{figure}
\begin{figure}
\epsfxsize=6truecm\epsfbox{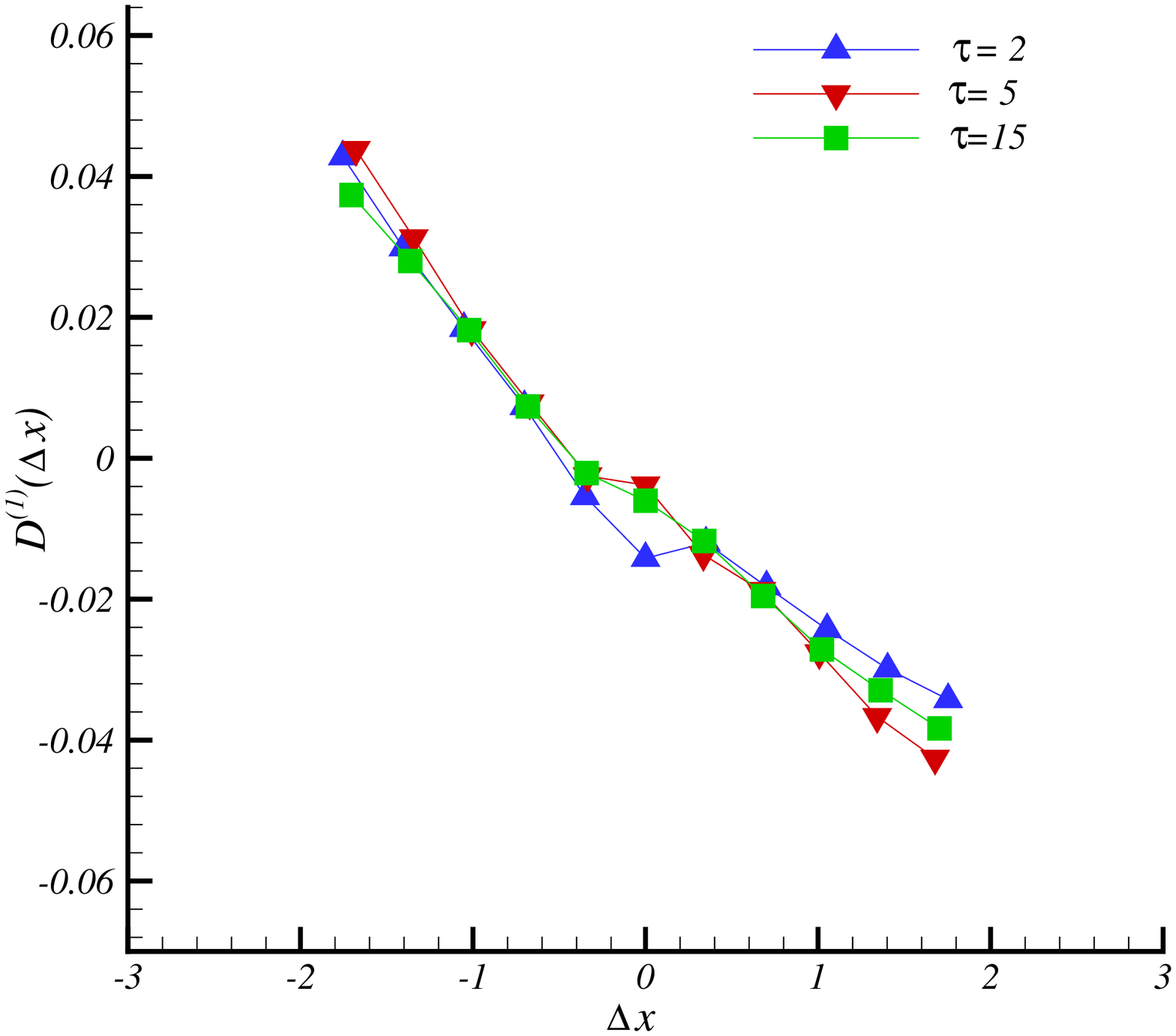}
\epsfxsize=6truecm\epsfbox{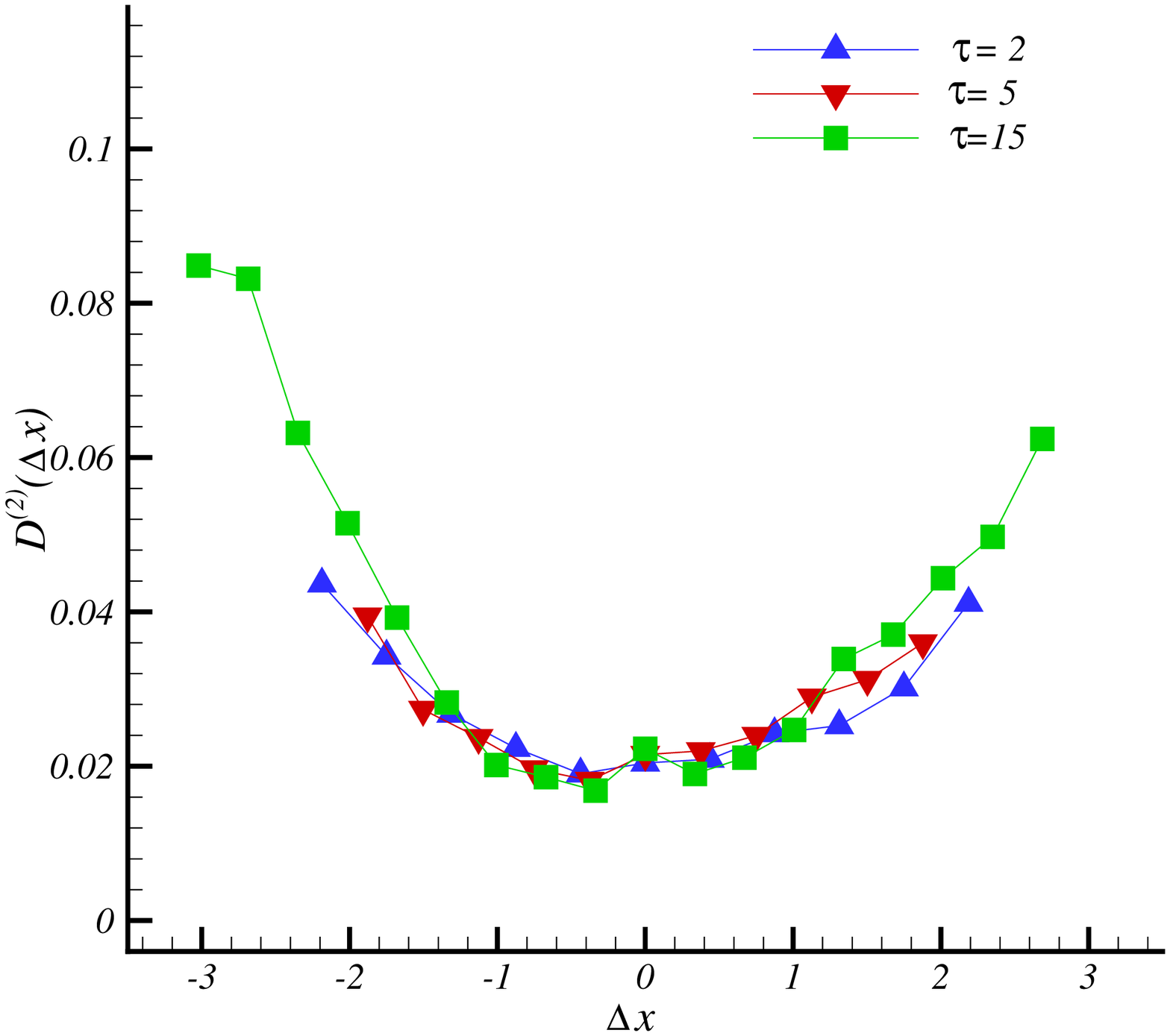} \narrowtext \caption{ The drift
and diffusion coefficients $D^{(1)}( \Delta x )$ and
$D^{(2)}(\Delta x)$ are estimated from the Eq. (3) for typical
patients with heart failure, and follow linear and quadratic
behavior, respectively.}
\end{figure}
\begin{figure}
\epsfxsize=7truecm\epsfbox{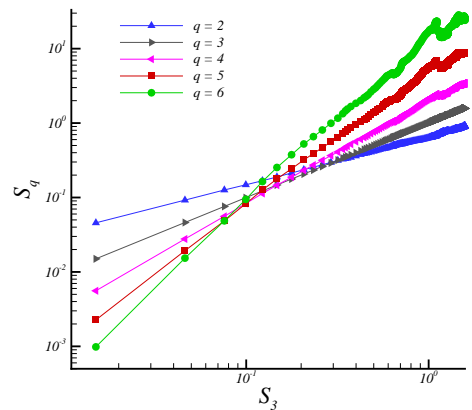} \narrowtext \caption{
Generalized scaling analysis of a typical healthy subject.
Structure functions $S_{q}$ are displayed versus $S_{3}$ in
log-log scale.}
 \end{figure}
In such cases, one computes the Kramers-Moyal (KM) coefficients
for the {\it increments} of interbeat intervals fluctuatations,
$\Delta x(\tau)=x(t+\tau)- x(t)$, rather than the time series
$x(t)$ itself. One then checks whether the first and second KM
coefficients that represent, respectively, the drift and diffusion
coefficients in a FP equation, have well-defined and finite
values, while the third- and fourth-order KM coefficients are
small. According to the Pawula`s theorem,$^{24}$ the KM expansion,
\begin{equation}
\frac{\partial}{\partial t}P(x,t|x_0,t_0)=\sum_{k=1}^\infty
\left(- \frac{\partial}{\partial
x}\right)^k\left[D^{(k)}(x,t)P(x,t|x_0,t_0)\right]\;,
\end{equation}
can be truncated after the second (diffusive) term, provided that
the third- and fourth-order coefficient $D^{(4)}$ vanish, or are
very small compared with the first two coefficients. If so, which
is often the case, then the KM expansion, Eq. (12), reduces to a
FP evolution equation. In that case, a FP equation is numerically
constructed by computing its drift and diffusion coefficients for
the PDF $P(\Delta x,\tau)$ which, in turn, is used to gain
information on evolution of the shape of the PDF as a function of
the time scale $\tau$. In essence, if the first two KM
coefficients are found to have numerically-meaningful values
(i.e., not very small), while the third and higher coefficients
are small compared with the first two coefficients, the above
reconstruction method - Eqs. (2)-(5) - are used for the {\it
increments} of the times series, rather than the time series
itself.

Therefore, carrying out the same type of computations described
above, but now for the increaments $\Delta x(t)$, the following
results are computed for the healthy subjects,$^{28}$
\begin{equation}
D^{(1)}(\Delta x,\tau)=-(3\Delta x+0.46)\times 10^{-2}\;,
\end{equation}
\begin{equation}
D^{(2)}(\Delta x,\tau)=\left[\left(1+11\tau^{-1}\right)(\Delta
x)^2+ \left(5.7+28.7\tau^{-1}\right)\right]\times 10^{-2}\;,
\end{equation}
whereas for the patients with CHF we obtain,
\begin{equation}
D^{(1)}(\Delta x,\tau)=-(1.3\Delta x+0.18)\times 10^{-2}\;,
\end{equation}
\begin{equation}
D^{(2)}(\Delta x,\tau)=\left[\left(5+5\tau^{-1}\right)(\Delta x)^2
+\left(13+66\tau^{-1}\right)\right]\times 10^{-3}\;.
\end{equation}
Estimates of the above coefficients are less accurate for large
values of $\Delta x$. Also computed are the {\it average} of the
coefficients $D^{(1)}$ and $D^{(2)}$ for the entire set of the
healthy subjects, as well as those with CHF. Moreover, $D^{(4)}$
is about $\frac{1}{10}D^{(2)}$ for the healthy subjects, and about
$\frac{1}{20}D^{(2)}$ for those with CHF. Therefore, the KM
expansion can indeed be truncated beyond the second term, and the
FP formulation is numerically justified.

Equations (13)-(16) state that the drift coefficients for the
healthy subjects and those with CHF have the same order of
magnitude, whereas the diffusion coefficients for the given $\tau$
and $\Delta x$ differ by about one order of magnitude. This points
to a relatively simple way of distinguishing the two classes of
the subjects. Moreover, the $\tau$-dependence of the diffusion
coefficient for the healthy subjects is stronger than that of
those with CHF (in the sense that the numerical coefficients of
the $\tau^{-1}$ are larger for the healthy subjects). These are
shown in Figures 7 and 8. Note also that these results are
consistent with those presented earlier for the time series $x(t)$
itself, in terms of distinguishing the two classes of patients
through their different drift and diffusion coefficients.

The strong $\tau-$dependence of the diffusion coefficient
$D^{(2)}$ for the healthy subjects indicates that the nature of
the PDF of their increments $\Delta x$ for given $\tau$, i.e.,
$P(\Delta x,\tau)$, is {\it intermittent}, and that its shape
should change strongly with $\tau$. However, for the patients with
CHF the PDF is not so sensitive to the change of the time scale
$\tau$, hence indicating that the increments' fluctuations for
these patients is {\it not} intermittent.

\vskip 2cm

 \noindent{\bf  The Extended Self-Similarity of Interbeat Intervals in Human
Subjects}

\vskip 2cm

Let's look at another computational method for distinguishing
healthy subjects from CHF patients. The method
is based on the concept of extended self-similarity (ESS) of a
time series. This concept is particularly useful if the time
series for interbeat fluctuations (or other types of time series)
do not, as is often the case, exhibit scaling over a broad
interval. In such cases, the time interval in which the structure
function of the time series, i.e.,
\begin{equation}
S_q(\tau)=\langle|x(t+\tau)-x(t)|^q\rangle\;,
\end{equation}
behaves as
\begin{equation}
S_q(\tau)\sim\tau^{\xi_q}\;,
\end{equation}
is small, in which case the existence of scale invariance in the
data can be questioned. However, instead of rejecting outright the
existence of scale invariance, one must first explore the
possibility of the data being scale invariance via the concept of
ESS.

The ESS is a powerful tool for checking non-Gaussian properties of
data,$^{31,32}$ and has been used extensively in research on
turbulent flows. Indeed, when analyzing the interbeat time series
for human subjects (and other types of time series), one can, in
addition to the $\tau$-dependence of the structure function,
compute a generalized form of scaling using the ESS concept. In
many cases, when the structure functions $S_q(\tau)$ are plotted
against a structure function of a specific order, say $S_3(\tau)$,
an extended scaling regime is found according to,$^{31,32}$
\begin{equation}
S_q(\tau)\sim S_3(\tau)^{\zeta_q}\;.
\end{equation}
Clearly, meaningful results are restricted to the regime where
$S_3$ is {\it monotonic.} For any Gaussian process the exponents
$\zeta_q$ follow a simple equation,
\begin{equation}
\zeta_q=\frac{1}{3}q\;.
\end{equation}
Therefore, systematic deviation from the simple scaling relation,
Eq. (20), should be interpreted as deviation from Gaussianity. An
additional remarkable property of the ESS is that it holds rather
well even in situations when the ordinary scaling does not exit,
or cannot be detected due to small scaling range, which is the
case for the data analyzed here.

Using the ESS concept, we analyzed$^{33}$ the fluctuations in
human heartbeat rates of healthy subjects and those with CHF,
analyzed earlier with the reconstruction method. The results are
shown in Figure 9. For a typical healthy subject an improved
scaling behavior of the time series is indicated by the ESS. In
Figure 10 the computed scaling exponents $\zeta_q$ of the
structure functions are plotted against the order $q$. A
mono-fractal time series corresponds to linear dependence of
$\zeta_q$ on $q$, whereas for a multifractal time series $\zeta_q$
depends nonlinearly on $q$. The constantly changing curvature of
the computed $\zeta_q$ for the healthy subjects suggests
multifractality of their corresponding time series. In contrast,
$\zeta_q$ is essentially linear for the patients with CHF,
indicating mono- or simple fractal behavior.

It is well-known that the moments with $q<1$ and $q>1$ are
related, respectively, to the frequent and rare events in the time
series.$^{30,31}$ Thus, for the data considered here one may also
be interested in the frequent events in the interbeats. In Figure
11 we show the results for the moment $q=0.1$ against a
third-order structure function for healthy subjects and those with
CHF. There are two interesting features in Figure 11. First, the
starting point of $S_{0.1}(\tau)$ versus $S_3(\tau)$ is different
for the data for healthy subjects and patients with CHF. To
determine the distance form the origin, we define,$^{34}$
\begin{equation}
T(\tau=1)=[S^2_{0.1}(\tau=1)+S^2_3(\tau=1)]^{1/2}\;.
\end{equation}
The second important feature of Figure 11 is that there is a
well-defined $\tau^*$ beyond which the plot of $S_{0.1}(\tau)$
versus $S_3(\tau)$ is multi-valued. One can estimate $\tau^*$ by
checking when $S_3(\tau)> S_3(\tau+1)$. Moreover, if we define
$T(\tau^*)$ by,
\begin{equation}
T(\tau^*)=[S^2_{0.1}(\tau^*)+S^2_3(\tau^*)]^{1/2}\;,
\end{equation}
then, there is a time scale $\tau^*$ such that the values of the
third moment before and after $\tau ^*$ are almost the same. Thus,
the quantity $\tau^*$ plays a role of a local mirror on the time
axis. In other words, locally, $S_3(\tau)$ for $\tau<\tau^*$ and
$\tau>\tau^*$ has almost the same value. In Figure 11, we show the
time scale $\tau^*$ and, therefore, indicate that values of
$T(\tau=1)$ and $T(\tau^*)$ of the interbeat fluctuations of the
healthy subjects and patients with CHF are {\it different}.

Table 1 presents the computed values of $T_1=T(\tau=1)$ for both
healthy subjects and those with CHF. To compute the results, we
first rescaled the data sets by their standard deviation and,
therefore, the $T_1$ values are dimensionless. The average value
of $T_1$ for the healthy subjects is, $\bar{T}(\tau=1)\simeq
0.5848$, with its standard deviations being, $\sigma \simeq
0.065$. The corresponding values for the daytime records of the
patients with CHF are, $\bar{T}(\tau=1)\simeq 0.5077$  and
$\sigma\simeq 0.03$, respectively. Therefore, on average, healthy
subjects possess $T_1$ values that are greater than those of the
patients with CHF. However, note that $T_1$ for the various data
sets do not have large enough differences to be able to
distinguish unambiguously the data sets. Indeed, as Table 1
indicates, three of the data sets (belonging to the day- and
nighttime records of one healthy subject, with the third one
belonging to the daytime of another healthy subject) overlap.

To develop a more definitive criterion for distinguishing the data
for various subjects, we compute values of $T(\tau^*)$. The
results are listed in Table 2. It is evident that, in this case,
there is no overlap in the data sets. Indeed, the $T(\tau^*)$
values for the healthy subjects are larger, by a factor of about
3, than those for the patients with CHF, hence providing an
unambiguous way of distinguishing the data sets for healthy
subjects and those with CHF.
\begin{figure}
\epsfxsize=7truecm\epsfbox{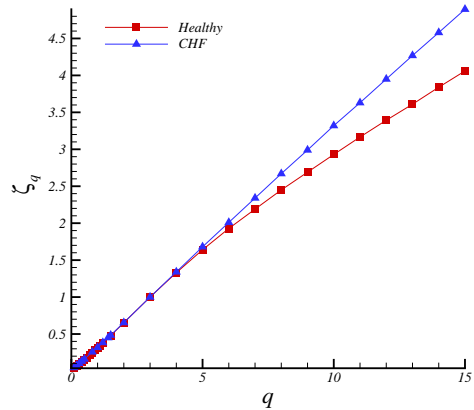} \narrowtext \caption{The plot of
$\zeta_{q}$ versus $q$, which has a linear dependence on $q$ for
CHF subjects, but a nonlinear dependence on $q$ for the healthy
subjects.}
 \end{figure}
\begin{figure}
\epsfxsize=7truecm\epsfbox{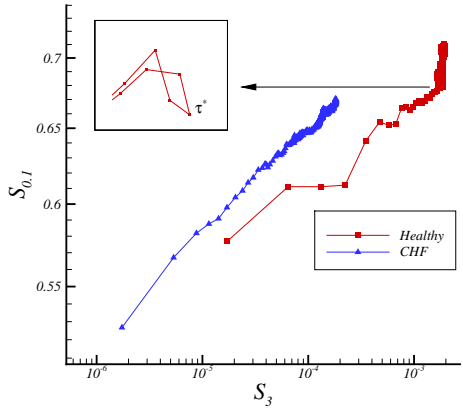} \narrowtext \caption{Plot of
$S_{0.1}$ against $S_3(\tau)$ for a healthy and one with CHF. The
results indicate that the starting points are different for
healthy subjects and those with CHF. Moreover, for the both data
set there is a well-defined $\tau^*$ at which $S_3(\tau) >
S_3(\tau+1)$.}
\end{figure}
\begin{table}[htb]
\begin{center}
\caption{\label{Tb2}Values of $T_1$ for the healthy subjects (day
time and night time ) and those with CHF.}

\medskip
\begin{tabular}{|c|c|}
\hline $Healthy$&$CHF$$\hspace{1.8cm}$\\\hline
$\hspace{1.8cm}$0.658$\hspace{1.8cm}$&0.557$\hspace{1.8cm}$\\\hline
$\hspace{1.8cm}$0.672$\hspace{1.8cm}$&0.565$\hspace{1.8cm}$\\\hline
$\hspace{1.8cm}$0.614$\hspace{1.8cm}$&0.539$\hspace{1.8cm}$\\\hline
$\hspace{1.8cm}$0.605$\hspace{1.8cm}$&0.526$\hspace{1.8cm}$\\\hline
$\hspace{1.8cm}$0.583$\hspace{1.8cm}$&0.512$\hspace{1.8cm}$\\\hline
$\hspace{1.8cm}$0.581$\hspace{1.8cm}$&0.493$\hspace{1.8cm}$\\\hline
$\hspace{1.8cm}$0.576$\hspace{1.8cm}$&0.492$\hspace{1.8cm}$\\\hline
$\hspace{1.8cm}$0.558$\hspace{1.8cm}$&0.481$\hspace{1.8cm}$\\\hline
$\hspace{1.8cm}$0.494$\hspace{1.8cm}$&0.469$\hspace{1.8cm}$\\\hline
$\hspace{1.8cm}$0.480$\hspace{1.8cm}$&0.443$\hspace{1.8cm}$\\\hline

\end{tabular}
\end{center}
\end{table}

\begin{table}[htb]
\begin{center}
\caption{\label{Tb2}Values of the $T(\tau^*)$ for the healthy
subjects (day time and night time) and for those with CHF.}

\medskip
\begin{tabular}{|c|c|}
\hline $Healthy$&$CHF$$\hspace{1.8cm}$\\\hline
$\hspace{1.8cm}$3.08$\hspace{1.8cm}$&0.741$\hspace{1.8cm}$\\\hline
$\hspace{1.8cm}$2.68$\hspace{1.8cm}$&0.714$\hspace{1.8cm}$\\\hline
$\hspace{1.8cm}$2.34$\hspace{1.8cm}$&0.685$\hspace{1.8cm}$\\\hline
$\hspace{1.8cm}$1.92$\hspace{1.8cm}$&0.681$\hspace{1.8cm}$\\\hline
$\hspace{1.8cm}$1.86$\hspace{1.8cm}$&0.675$\hspace{1.8cm}$\\\hline
$\hspace{1.8cm}$1.42$\hspace{1.8cm}$&0.632$\hspace{1.8cm}$\\\hline
$\hspace{1.8cm}$1.40$\hspace{1.8cm}$&0.573$\hspace{1.8cm}$\\\hline
$\hspace{1.8cm}$1.22$\hspace{1.8cm}$&0.728$\hspace{1.8cm}$\\\hline
$\hspace{1.8cm}$1.20$\hspace{1.8cm}$&0.552$\hspace{1.8cm}$\\\hline
$\hspace{1.8cm}$1.15$\hspace{1.8cm}$&0.465$\hspace{1.8cm}$\\\hline

\end{tabular}
\end{center}
\end{table}

\vskip 5cm

\noindent{\bf Comparison with other Methods}

\bigskip
\noindent

Stanley and colleagues$^{5,13,19,26,27,38}$ and other$^{35-37}$
analyzed the type of data that we consider in this review by
different methods. Their analyses indicate that there may be
long-range correlations in the data, characterized by self-affine
fractal distributions, such as the fractional Brownian motion, the
power spectrum of which is given by,
\begin{equation}
S(f)\sim f^{-(2H+1)}\;,
\end{equation}
where $H$ is the Hurst exponent that characterizes the type of the
correlations in the data. Thus, healthy subjects are distinguished
from those with CHF in terms of the type of correlations that
might exist in the data: negative or antipersistent correlations
in the increments for $H<1/2$, as opposed to positive or
persistent correlations for $H>1/2$, and Brownian motion for $H=
1/2$. While this is an important and interesting result, it can
also be ambiguous and not very precise. For example, suppose that
the analysis of two time series yields two values of $H$, one
slightly larger and the second one slightly smaller than $1/2$.
Then, 2. It's thus difficult to state with confidence
that the two times series are really distinct.

The reconstruction method described earlier analyzes the
data in terms of the Markov processes' properties. As a result,
it distinguishes the data for healthy subjects from those
with CHF in terms of the differences between an FP equation's
drift and diffusion coefficients. Such differences are typically very significant and,
therefore, provide, in our view, an unambiguous way of
understanding the differences between the two groups of the
subjects, including for those series for which the Hurst exponents
are only slightly different. In addition, the computational
approach described in this review provides an unambiguous way of
{\it reconstructing} the data, hence providing a means of {\it
predicting} the behavior of the data over periods of time that are
on the order of their Markov time scales.

Although it remains to be tested, we believe that, together, all
the computational methods that have been described in this review
are more sensitive to small differences between the data for the
two groups of the subjects and, therefore, might eventually
provide a diagnostic tool for {\it early} detection of CHF in
patients.

Finally, the computational approaches described in this review are
quite general, and may be used for analyzing times series that
represent the dynamics of completely unrelated phenomena. For
example, we have used the concepts of Markov processes and
extended self-similarity to develop$^{34}$ a method for providing
short-term alerts for moderate and large earthquakes, as well as
making predictions for the price of oil.$^{39}$

\vskip 2cm

 \noindent{\bf Biographies:}

Mohammad Reza Rahimi Tabar is Associate Professor of Physics at
the Sharif University in Tehran, Iran. His main research areas are
conformal field theory, disordered systems, statistical theory of
turbulence, stochastic processes, seismic activity, and wave
localization. He can be reached at rahimitabar$@$sharif.edu.

Fatemeh Ghasemi is a post-doctoral fellow at Institute for Studies
in Theoretical Physics and Mathematics in Tehran, Iran. Her main
research interests are stochastic processes, analysis of time
series and of rough
surfaces, and seismic activity. She can be reached at \\
f$\_$ghasemi$@$mehr.sharif.edu.

Joachim Peinke is a Professor at the University of Oldenburg in
Germany. His main research areas are turbulence, econophysics,
pattern formation in nematic liquid crystals, stochastic
processes, and development of various sensors. He can be reached
at peinke$@$uni-oldenburg.de.

Rudolf Friedrich is a Professor at the University of M\"unster,
Germany. His main research areas are turbulence, econophysics,
pattern formation in nematic liquid crystals, and stochastic
processes. He can be reached at fiddir$@$uni-muenster.de .

Kamran Kaviani is Assistant Professor of Physics at Alzahra
University, in Tehran, Iran. His main research are stochastic
processes and field theory. He can be reached at
kaviani$@$yahoo.com.

Fatemeh Taghavi is a Ph.D. student in physics at Iran University
of Science and Technology in Tehran, Iran. Her main research areas
are stochastic processes and field theory. She can be reached at
f$\_$taghavi$@$iust.ac.ir.

Sara Sadeghi is a Ph.D. student in physics at Sharif University in
Tehran, Iran. Her main reseach area is analysis of complex
systems. She can be reached at s$\_$sara2001@yahoo.com.

Golnoosh Bizhani is pursuing her M.Sc. degree in physics at Sharif
University in Tehran, Iran. Her main reseach area is analysis of
complex systems. She can be reached at\\
golnoosh$\_$bizhani$@$yahoo.com.

Muhammad Sahimi is Professor of Chemical Engineering and Materials
Science, and NIOC Professor of Petroleum Engineering at the
University of Southern California in Los Angeles. His main
research areas are atomistic simulation of nanostructured
materials, modeling of large-scale porous media, wave propagation
in heterogeneous media, and analysis of stochastic processes. He
can be reached at moe$@$iran.usc.edu.

\end{document}